# Subclustering and Star Formation Efficiency in Three Protoclusters in the Central Molecular Zone

Suinan Zhang (张遂楠),[1] Xing Lu (吕行),[1] Adam Ginsburg,[2] Nazar Budaiev,[2] Yu Cheng,[3] Hauyu Baobab Liu,[4,5] Tie Liu,[1] Qizhou Zhang,[6] Keping Qiu,[7,8] Siyi Feng,[9] Thushara Pillai,[10] Xindi Tang,[11,12,13,14] Elisabeth A. C. Mills,[15] Qiuyi Luo,[1] Shanghuo Li,[7,8] Namitha Issac,[1] Xunchuan Liu,[1] Fengwei Xu,[16,17,18] Jennifer Wallace,[19] Xiaofeng Mai,[1] Yan-Kun Zhang,[1] Cara Battersby,[19] Steven N. Longmore,[20,21] and Zhiqiang Shen[1]

[1]*Shanghai Astronomical Observatory, Chinese Academy of Sciences, 80 Nandan Road, Shanghai 200030, P. R. China*
[2]*Department of Astronomy, University of Florida, P.O. Box 112055, Gainesville, FL 32611, USA*
[3]*National Astronomical Observatory of Japan, 2-21-1 Osawa, Mitaka, Tokyo 181-8588, Japan*
[4]*Department of Physics, National Sun Yat-Sen University, No. 70, Lien-Hai Road, Kaohsiung City 80424, Taiwan, R.O.C.*
[5]*Center of Astronomy and Gravitation, National Taiwan Normal University, Taipei 116, Taiwan*
[6]*Center for Astrophysics | Harvard & Smithsonian, 60 Garden Street, Cambridge, MA 02138, USA*
[7]*School of Astronomy and Space Science, Nanjing University, 163 Xianlin Avenue, Nanjing 210023, Peoples Republic of China*
[8]*Key Laboratory of Modern Astronomy and Astrophysics (Nanjing University), Ministry of Education, Nanjing 210023, Peoples Republic of China*
[9]*Department of Astronomy, Xiamen University, Zengcuo'an West Road, Xiamen, 361005, Peoples Republic of China*
[10]*Haystack Observatory, Massachusetts Institute of Technology, 99 Millstone Road, Westford, MA 01886, USA*
[11]*Xinjiang Astronomical Observatory, 150 Science 1-Street, Urumqi, Xinjiang 830011, Peoples Republic of China*
[12]*University of Chinese Academy of Sciences, Beijing, 100080, Peoples Republic of China*
[13]*Key Laboratory of Radio Astronomy,Chinese Academy of Sciences, Urumqi, 830011, Peoples Republic of China*
[14]*Xinjiang Key Laboratory of Radio Astrophysics,Urumqi, 830011, Peoples Republic of China*
[15]*Department of Physics and Astronomy, University of Kansas, 1251 Wescoe Hall Drive, Lawrence, KS 66045, USA*
[16]*I. Physikalisches Institut, Universität zu Köln, Zülpicher Str. 77, D-50937 Köln, Germany*
[17]*Kavli Institute for Astronomy and Astrophysics, Peking University, Beijing 100871, People's Republic of China*
[18]*Department of Astronomy, School of Physics, Peking University, Beijing, 100871, People's Republic of China*
[19]*Department of Physics, University of Connecticut, 196A Auditorium Road, Storrs, CT 06269, USA*
[20]*Astrophysics Research Institute, Liverpool John Moores University, IC2, Liverpool Science Park, 146 Brownlow Hill, Liverpool, L3 5RF, United Kingdom*
[21]*Cosmic Origins Of Life (COOL) Research DAO, coolresearch.io*

## ABSTRACT

We present so far the highest resolution ($\sim 0\rlap{.}{''}04$) ALMA 1.3 mm continuum observations of three massive star-forming clumps in the Central Molecular Zone, namely 20 km s$^{-1}$ C1, 20 km s$^{-1}$ C4, and Sgr C C4, which reveal prevalent compact millimeter emission. We extract the compact emission with *astrodendro* and identify a total of 199 fragments with a typical size of $\sim$370 AU, which represent the first sample of candidates of protostellar envelopes and disks and kernels of prestellar cores in these clumps that are likely forming star clusters. Compared with the protoclusters in the Galactic disk, the three protoclusters display a higher level of hierarchical clustering, likely a result of the stronger turbulence in the CMZ clumps. Compared with the mini-starbursts in the CMZ, Sgr B2 M and N, the three protoclusters also show stronger subclustering in conjunction with a lack of massive fragments. The efficiency of high-mass star formation of the three protoclusters is on average one order of magnitude lower than that of Sgr B2 M and N, despite a similar overall efficiency of converting gas into stars. The lower efficiency of high-mass star formation in the three protoclusters is likely attributed to hierarchical cluster formation.

Corresponding author: Suinan Zhang

suinan.zhang@gmail.com

## 1. INTRODUCTION

The Central Molecular Zone (CMZ) is the innermost ∼500 pc of the Milky Way enriched with molecular gas of ∼2–6 ×$10^7$ $M_\odot$ (Ferrière et al. 2007). Observations over the past decades have revealed inefficient star formation in the CMZ with an overall star formation rate (SFR) of ∼0.07 $M_\odot$ yr$^{-1}$, which is about one order of magnitude lower than that predicted by the dense gas-star formation relation (Henshaw et al. 2023). This has been attributed to the extreme physical conditions in the CMZ, characterized by high gas temperatures (≳50–100 K; Ao et al. 2013; Ginsburg et al. 2016), strong magnetic fields (≳1 mG; Pillai et al. 2015; Pan et al. 2024), and strong turbulence (Mach numbers ∼20–30 in pc-scale clouds; Shetty et al. 2012; Liu et al. 2013; Henshaw et al. 2016). However, due to limited angular resolutions and sensitivities, most studies investigate the inefficient star formation at cloud scale (∼1–10 pc; Kauffmann et al. 2017a,b; Battersby et al. 2024a,b).

Recent high-resolution (2000–5000 AU) ALMA observations have spatially resolved massive clouds in the CMZ, including the Sgr B2 cloud, Dust Ridge clouds, the 20 km s$^{-1}$ cloud, and the Sgr C cloud, into prestellar or protostellar cores (Ginsburg et al. 2018; Barnes et al. 2019; Lu et al. 2020, 2021; Williams et al. 2022). Among them, only Sgr B2 has been imaged with sufficiently high resolution (500–700 AU) to resolve individual young stellar objects (YSOs) (Budaiev et al. 2024). The authors have characterized the Class 0/I YSOs in Sgr B2 M and N, enabling star-counting-based SFR estimates.

Despite these progresses, it remains poorly explored if and how the extreme physical conditions in the CMZ affect the formation path of star clusters, and ultimately the star formation efficiency (SFE). To systematically characterize the star and cluster formation in the CMZ at the scale of individual forming stars (a few 100 AU, Beuther et al. 2018), we have carried out ALMA Band 6 observations of three massive star-forming clumps in the 20 km s$^{-1}$ cloud and Sgr C, namely 20 km s$^{-1}$ C1, 20 km s$^{-1}$ C4, and Sgr C C4 (Lu et al. 2019a), at an angular resolution of ∼0″.04 (equivalent to ∼330 AU at a distance of 8.277 kpc to the Galactic Center, GRAVITY Collaboration et al. 2022). The clumps are selected because they have strong turbulence (FWHM∼5–10 km s$^{-1}$, Krieger et al. 2017) and host ongoing high-mass star formation traced by masers, outflows, and ultra-compact H II regions (Lu et al. 2019a,b, 2021). They have typical sizes of ∼1 pc and gas masses of a few ×$10^3$ $M_\odot$ (Lu et al. 2019a).

In this Letter, we present so far the highest resolution ALMA 1.3 mm continuum data of the three clumps. We investigate the fragmentation and clustering for each clump and compare our sample with protoclusters in the Galactic disk and mini-starbursts in the CMZ. We estimate the SFR and SFE of our sample and combine the SFE with their clustering properties to discuss the implications to the cluster formation in the CMZ.

## 2. OBSERVATIONS

We observed the three clumps with the ALMA 12 m array at Band 6 (1.3 mm wavelength, with a central frequency at 225.987 GHz) in Cycle 6 (project ID: 2018.1.00641.S; PI: Xing Lu). The observations were taken between June 5 and July 15, 2019. The phase centers (J2000) of the three single-pointing observations were ($17^h45^m37^s.5528$, $-29°03'49''.365$), ($17^h45^m37^s.5704$, $-29°05'44''.232$), and ($17^h44^m40^s.3677$, $-29°28'14''.794$) for 20 km s$^{-1}$ C1, 20 km s$^{-1}$ C4, and Sgr C C4, respectively. The full width half maximum (FWHM) size of the primary beam at 1.3 mm is ∼24″.4. These observations were taken under the C-8 configuration with 42–48 antennas, resulting in a range of baselines of 82–15238 m. The Maximum Recoverable Scale (MRS) is ∼0″.65, which is equivalent to ∼5400 AU. The total on-source time is 64.1, 65.3, and 64.6 minutes for 20 km s$^{-1}$ C1, 20 km s$^{-1}$ C4, and Sgr C C4, respectively. J1924−2914 was observed for absolute flux and bandpass calibration with J1744−3116 for phase calibration.

The data were calibrated using the pipeline in the Common Astronomy Software Applications package (CASA) 5.4.0, and then self-calibrated and imaged using CASA 6.5.3. Line-free channels identified in the spectra of the visibilities were used to image the continuum with the *tclean* task. Two rounds of phase-only self-calibration were performed for 20 km s$^{-1}$ C1 and 20 km s$^{-1}$ C4. For Sgr C C4 where the peak intensity is up to 12.5 mJy beam$^{-1}$, we performed six rounds of phase-only and one round of amplitude self-calibration. The amplitude self-calibration results in a ≲ 4% change in the flux density. The self-calibrated data were imaged with the multiscale, multifrequency synthesis algorithm implemented in *tclean*, with a pixel size of 0″.007 and scales of [0,5,15,50] times the pixel size. The Briggs weighting scheme was chosen with a robust parameter of 0.5. The resulting synthesized beam is 0″.043 × 0″.029 with a position angle of 71.62°, equivalent to a linear scale of 360 AU × 240 AU. The 1$\sigma$ rms noise before the primary beam correction is 13 $\mu$Jy beam$^{-1}$ and 17 $\mu$Jy beam$^{-1}$ for 20 km s$^{-1}$ and Sgr C, respectively. The continuum images are available at 10.5281/zenodo.14767274.

## 3. RESULTS

Figure 1 shows the ALMA 1.3 mm continuum images in brightness temperatures. A rich population of compact structures distribute throughout each pc-scale clump. Figure 1 A−F provide zoom-in views of the selected regions in the left column, showcasing subclustering of compact structures within a few thousands AU.

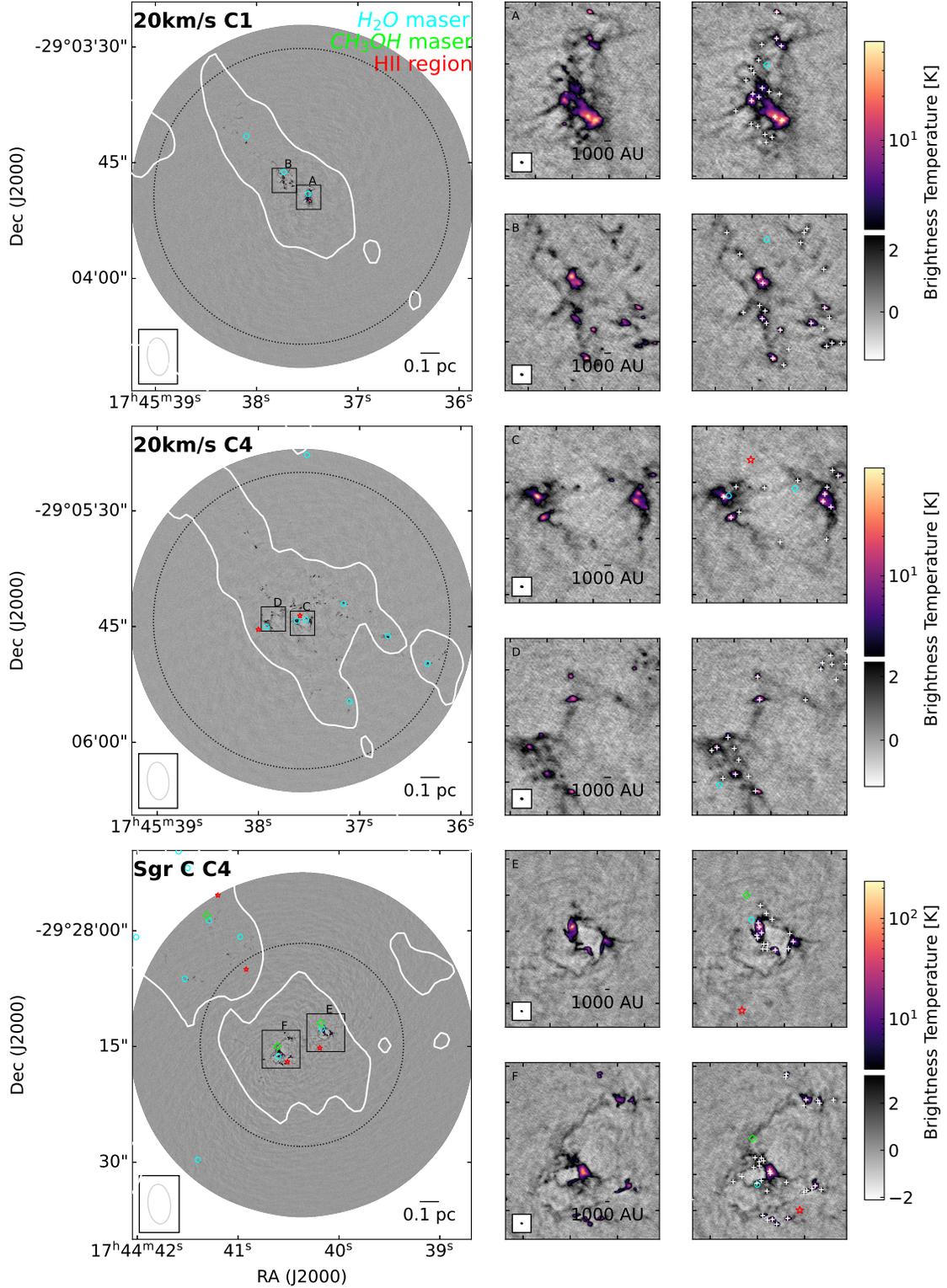

**Figure 1**: Brightness temperature images overlaid with white contours outlining the SMA 1.3 mm continuum from Lu et al. (2019a) at the level of $3\sigma_{\rm SMA}^{1.3\rm mm}$, where $\sigma_{\rm SMA}^{1.3\rm mm} = 3$ mJy beam$^{-1}$. Values lower than $10\sigma$ are shown in a linear greyscale, while those higher than $10\sigma$ are highlighted in a logarithmic colorscale. The locations of H$_2$O masers, Class II CH$_3$OH masers, and H II regions are marked in cyan, green, and red (Lu et al. 2019a,b). The black dotted circles indicate 20% of the primary beam response for 20 km s$^{-1}$ C1 and C4 and 50% for Sgr C C4. The right panels are the zoomed-in views of subregions outlined by the black squares in the left panels. The squares are labeled from "A" to "F", with a size of $\sim 3''$ for A, B, C, D and $\sim 5''$ for E and F. The crosses mark the locations of the fragments in the "robust" catalog. The synthesized beams of the SMA 1.3 mm continuum and the ALMA 1.3 mm continuum are shown in the bottom left corners of the left and right panels, respectively.

### 3.1. *Extraction of compact structures*

We extract compact structures in the 1.3 mm continuum images using the *astrodendro* package[1]. The structures are first identified in the images before primary beam correction that have a uniform noise level. The subsequent analyses are performed with the primary-beam-corrected images.

We adopt a minimum value of $5\sigma$, a minimum delta of $3\sigma$, and a minimum number of pixels in each *astrodendro* leaf that equals to the number of pixels per synthesized beam to construct a "robust" catalog. In addition, we employ a relaxed set of parameters of *astrodendro* (a minimum value of $3\sigma$, a minimum delta of $2\sigma$) to generate a "complete" catalog. We visually identify the boundaries of clusters based on the grouping of the ∼0.1-pc scale SMA cores (Lu et al. 2019a). The sources outside a chosen level of the primary beam response of our ALMA data (20% for 20 km s$^{-1}$ C1 and C4, and 50% for Sgr C C4) are discarded, as they are embedded in the SMA cores of other groups that are not fully covered by our ALMA observations.

A total of 199 sources are identified, with 52 in 20 km s$^{-1}$ C1, 97 in 20 km s$^{-1}$ C4, and 50 in Sgr C C4 for the "robust" catalog. For the "complete" catalog, the number of identified sources is 382, with 100 in 20 km s$^{-1}$ C1, 176 in 20 km s$^{-1}$ C4, and 106 in Sgr C C4. The short version of the "robust" catalog is provided in Table 1. The full catalogs are available in machine-readable forms in the online journal. The "robust" catalog is 90% complete at $57\sigma$ level (equivalent to 0.34 $M_\odot$ with our assumptions in Section 3.2) for 20 km s$^{-1}$ C1, $57\sigma$ (0.34 $M_\odot$) for 20 km s$^{-1}$ C4, and $26\sigma$ (0.20 $M_\odot$) for Sgr C C4. We employ the "robust" catalog for all the analyses presented in this letter.

We note that these identified sources of the size of $\gtrsim$300 AU differ markedly from the ∼0.1 pc "cores" resolved by the SMA observations at an angular resolution of ∼4$\farcs$0 (Lu et al. 2019a, see Figure 1) and the ∼0.01 pc "condensations" by previous ALMA observations at a resolution of 0$\farcs$2 (Lu et al. 2020) regarding spatial scales. We refer to these sources as "fragments" hereafter. Comparing the spatial distributions of the fragments and the cores and condensations at larger scales, we find that most fragments are spatially associated with those larger-scale structures, indicating multi-scale fragmentation within these molecular clouds. A comprehensive discussion of multi-scale fragmentation is beyond the scope of this letter and will be presented in a separate publication.

In particular, we evaluate the gravitational boundness of each group of the ∼0.1 pc cores using the virial parameter defined by $\alpha = \frac{5\sigma_{\text{clump}}^2 R_{\text{eff}}}{GM}$ (Bertoldi & McKee 1992). The masses and centroid velocities of cores are adopted from (Lu et al. 2019a). The centroid velocity of each core group is taken to be the mean centroid velocity of the cores. The velocity dispersion of each core group, $\sigma_{\text{clump}}$, is taken to be the standard deviation of centroid velocities of the cores. This yields an $\alpha$ of <1 for all three clumps, suggesting that the parental structures of fragments are gravitationally bound to each other and the fragments embedded within are likely forming star clusters. Therefore, we refer to the three clumps as protoclusters.

### 3.2. *Physical properties of fragments*

30% of the fragments have peak brightness temperatures higher than 10 K, the typical temperature at the center of prestellar cores without ongoing star formation (Pagani et al. 2007; Launhardt et al. 2013). 16% of the fragments have peak brightness temperatures >20 K, the typical dust temperature of the 20 km s$^{-1}$ cloud and the Sgr C cloud (Battersby et al. 2024a), indicating internal heating from protostars. The effective radius $R_e$ of the fragments is defined as $R_e = (A/\pi)^{1/2}$, where $A$ is the fragment area reported by *dendrogram*. $R_e$ has a median value of 370 AU, as shown in Figure 2(a), which is close to the resolution of 330 AU, suggesting marginally resolved compact structures. Therefore, the fragments with the compact 1.3 mm continuum emission likely internally heated represent candidates of protostellar envelopes and disks, with the rest candidates of the kernels of prestellar cores (e.g., Hirano et al. 2024).

We derive the physical properties of the fragments based on their 1.3 mm continuum emission. As the exact optical depths are unconstrained by current observations, the 1.3 mm continuum emission is assumed to be optically thin in these calculations, which are lower limits for the fragment masses and densities. The gas masses are calculated following

$$M_{\text{gas}} = \frac{S_\nu D^2 R}{B_\nu(T_{\text{dust}})\kappa_\nu}, \tag{1}$$

where $S_\nu$ is the flux density, $D$ is the distance, $R$ is the gas-to-dust mass ratio which is assumed to be 100, $B_\nu(T_{\text{dust}})$ is the Planck function for the dust temperature $T_{\text{dust}}$ that is assumed to be 50 K for all fragments, and $\kappa_\nu$ is the dust opacity that is adopted to be $\kappa_\nu = 1.11$ cm$^2$ g$^{-1}$ for an MRN distribution with thin ice mantles after $10^5$ years of coagulation at $10^8$ cm$^{-3}$ (Ossenkopf & Henning 1994). The fragment density $n(\text{H}_2)$ is then derived with the assumption of a uniform sphere following $n(\text{H}_2) = M_{\text{gas}}/(4\pi R_e^3/3)/(\mu_{\text{H}_2} m_\text{H})$, where the mean molecular weight per hydrogen molecule $\mu_{\text{H}_2}$ is adopted to be 2.8 (Kauffmann et al. 2008). Figure 2(a) shows the distribution of masses and effective radii of the fragments. The fragments have a median gas mass of 0.36 $M_\odot$ and a median gas density of $2.2 \times 10^8$ cm$^{-3}$.

Uncertainties in mass estimates mainly arise from the assumptions of optically thin dust emission and uniform dust temperature, choices of dust opacity, and potential contamination of free-free emission. A lower limit of the optical depth can be determined by $\tau_\nu = \kappa_\nu \Sigma$, where $\Sigma$ is the dust surface density calculated under the optically thin as-sumption. This method yields a $\tau_\nu$ of up to 2 and 9 for 20 km s$^{-1}$ C1 and Sgr C C4, respectively, and $\tau_\nu < 1$ for 20 km s$^{-1}$ C4. This suggests that these fragments are at least partially optically thick. In addition, for fragments internally heated by protostars, their kinetic temperatures could be up to >100 K (e.g., Zhang et al. 2025), making their masses overestimated by a factor of a few. This will be further discussed in Appendix Section A.1. Also, different choices of $\kappa_\nu$ for an MRN distribution with thin or thick ice mantles at $10^6$–$10^8$ cm$^{-3}$ could result in ≲ 20% difference in mass estimates. Lastly, at 1.3 mm, the thermal dust emission might be contaminated by free-free emission from unresolved H II regions. This effect remains to be further verified with multi-wavelength high-resolution observations.

---

[1] http://www.dendrograms.org/

**Table 1**: The "robust" catalog.

| Clump | ID | R.A., Decl. (J2000) | Radius (AU) | Flux (mJy) | Peak Brightness Temperature (K) | Gas Mass ($M_\odot$) | Gas Density ($cm^{-3}$) |
|---|---|---|---|---|---|---|---|
| 20 km s$^{-1}$ C1 | 1 | 17:45:37.49, −29:03:50.40 | 340 | 0.32 | 2.2 | 0.14 | $1.04 \times 10^8$ |
| 20 km s$^{-1}$ C1 | 2 | 17:45:37.48, −29:03:50.31 | 480 | 0.88 | 2.7 | 0.37 | $1.03 \times 10^8$ |
| 20 km s$^{-1}$ C1 | 3 | 17:45:37.49, −29:03:50.26 | 350 | 0.40 | 2.6 | 0.17 | $1.22 \times 10^8$ |
| 20 km s$^{-1}$ C1 | 4 | 17:45:37.50, −29:03:50.11 | 310 | 0.31 | 2.9 | 0.13 | $1.35 \times 10^8$ |
| 20 km s$^{-1}$ C1 | 5 | 17:45:37.51, −29:03:50.07 | 350 | 0.67 | 4.5 | 0.28 | $2.05 \times 10^8$ |
| 20 km s$^{-1}$ C1 | 6 | 17:45:37.47, −29:03:50.01 | 460 | 9.73 | 31.9 | 4.09 | $1.27 \times 10^9$ |
| 20 km s$^{-1}$ C1 | 7 | 17:45:37.48, −29:03:49.95 | 400 | 7.61 | 34.0 | 3.20 | $1.47 \times 10^9$ |
| 20 km s$^{-1}$ C1 | 8 | 17:45:37.53, −29:03:49.84 | 350 | 0.37 | 2.3 | 0.16 | $1.07 \times 10^9$ |
| 20 km s$^{-1}$ C1 | 9 | 17:45:37.51, −29:03:49.66 | 530 | 4.49 | 10.8 | 1.89 | $3.92 \times 10^8$ |
| 20 km s$^{-1}$ C1 | 10 | 17:45:37.47, −29:03:49.54 | 570 | 0.84 | 1.7 | 0.35 | $5.74 \times 10^7$ |

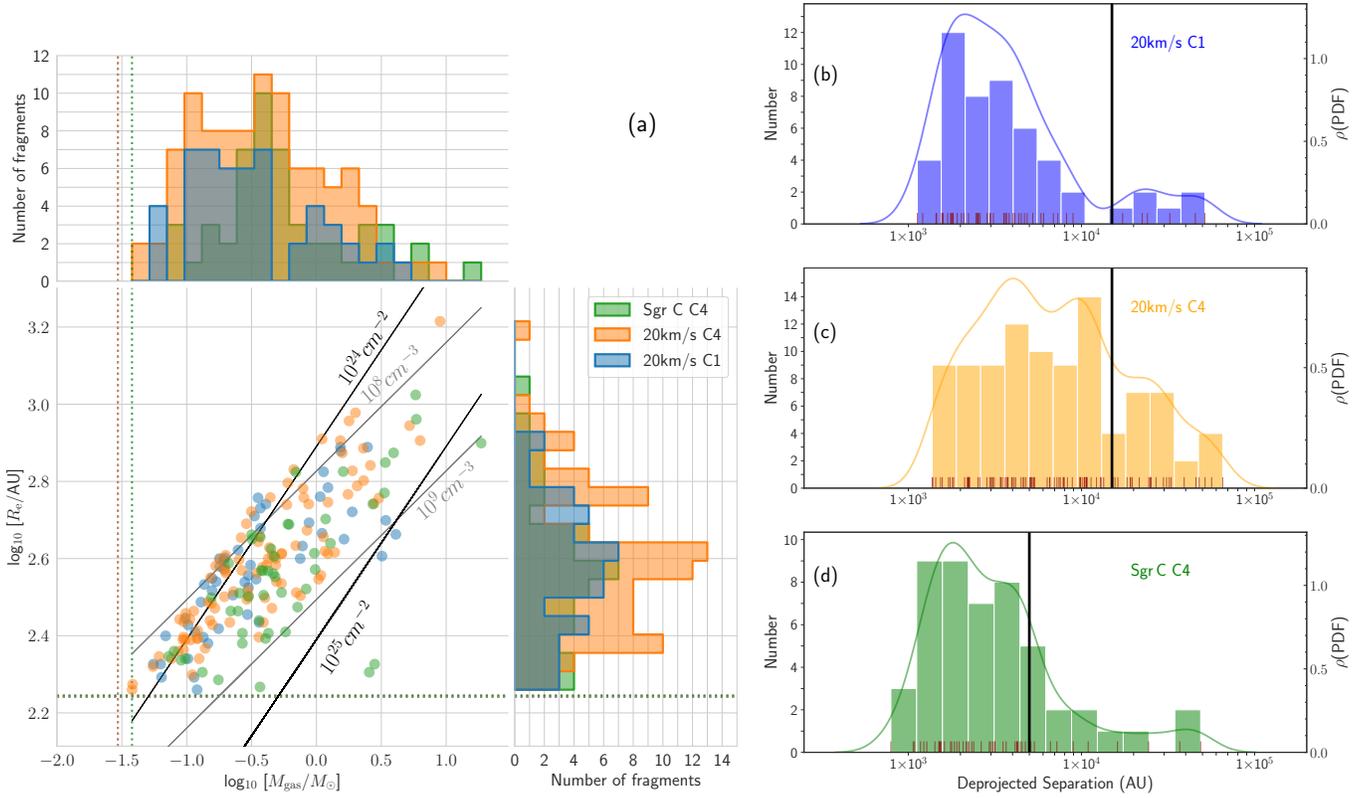

**Figure 2**: (a): Distribution of physical properties. The main plot shows the radii versus masses of the fragments. The marginal plots show the histograms of radii and masses on the top and right panels, respectively. The fragments in different clumps are color-coded. The vertical dotted lines mark masses corresponding to $5\sigma$. The horizontal dotted lines show effective radii of the synthesized beams. The black solid lines show constant column densities of $10^{24}$ and $10^{25}$ cm$^{-2}$. The grey solid lines show constant number densities of $10^8$ and $10^9$ cm$^{-3}$. (b)–(d): Distribution of deprojected separations of the three clumps. Each panel shows deprojected separations in red sticks along with the corresponding histogram and the kernel density estimate (KDE). The vertical lines represent the thermal Jeans length.

## 4. DISCUSSION

### 4.1. *Separations between fragments smaller than the thermal Jeans length*

To characterize the separations between the fragments, we apply the Minimum Spanning Tree (MST) technique (Barrow et al. 1985) implemented in the *FragMent* package[2] (Clarke et al. 2019). The MST method finds a way to connect the fragments which gives a minimum total length without any closed loop. We calculate the deprojected separations between fragments by dividing the separations determined with MST over a factor of $2/\pi$ (Sanhueza et al. 2019) and show the results in Figure 2(b)–(d).

To investigate the physical mechanism of fragment formation, we estimate the thermal Jeans length of the 0.1 pc cores following

$$\lambda_{\rm J} = c_s (\frac{\pi}{G\rho})^{1/2}, \quad (2)$$

where $c_s$ is the sound speed and $\rho$ is the gas density. While the gas and dust are collisionally well coupled at high densities, it has been observed that the gas temperature is higher than the dust temperature at the cloud scale in the CMZ due to turbulent heating (Ao et al. 2013; Ginsburg et al. 2016; Battersby et al. 2024a). With a gas temperature of $T_{\rm gas} = 100$ K (Lu et al. 2017) and an average gas density of the cores (Lu et al. 2019a), the thermal Jeans length is estimated to be ∼15000 AU for the two clumps in the 20 km s$^{-1}$ cloud and ∼5000 AU for Sgr C C4. These values are larger than the typical deprojected separations between fragments (see Figure 2(b)–(d)).

Several possibilities could explain the observed small separations. i) The thermal Jeans lengths might be overestimated by adopting a mean gas density whereas the volume filling factor may be smaller than 1. ii) Other than Jeans fragmentation, disk fragmentation might also contribute at these small scales. In particular, we have identified fragments in a massive Keplerian disk candidate surrounding a proto-O star with $M_* \sim 32\ M_\odot$ (see region E in Figure 1; Lu et al. 2022), which might be a result of disk fragmentation. iii) The separations between fragments could also be influenced by dynamical processes such as migration, ejection due to N-body interactions, and gravitational capture (Offner et al. 2023). Distinguishing between these mechanisms requires a thorough investigation of the fragment kinematics using spectral line emission and will be done in future work.

### 4.2. *Subclustering of the protoclusters*

We employ the MST technique and the structure parameter $\mathcal{Q}$ to characterize the clustering of the three protoclusters following Cartwright & Whitworth (2004). The $\mathcal{Q}$ parameter is defined by

$$\mathcal{Q} = \frac{\bar{m}}{\bar{s}}. \quad (3)$$

The $\bar{m}$ is the normalized mean edge length of the MST defined by

$$\bar{m} = \sum_{i=1}^{N-1} \frac{L_i}{(NA)^{1/2}}, \quad (4)$$

where $N$ is the number of fragments, $L_i$ is each edge length, and $A$ is the cluster area (given by $A = \pi R_c^2$ with $R_c$ being the distance between the mean position of all fragments and the most distant fragment). $\bar{s}$ is the normalized correlation length defined by

$$\bar{s} = \frac{L_{av}}{R_c}, \quad (5)$$

where $L_{av}$ is the average length of all the edges. A $\mathcal{Q}$ value of >0.8 suggests a centrally condensed configuration, while $\mathcal{Q} < 0.8$ suggests hierarchical clustering. The $\mathcal{Q}$ values calculated for the three protoclusters are smaller than 0.8, with $0.58 \pm 0.02$ for 20 km s$^{-1}$ C1, $0.56 \pm 0.02$ for 20 km s$^{-1}$ C4, and $0.45 \pm 0.01$ for Sgr C C4 (see uncertainty estimation in Section A.2). These values suggest that the three protoclusters are strongly subclustered, in agreement with the visual inspection for Figure 1.

Applying similar methods to high-mass star-forming regions in the Galactic disk, Zhang et al. (2022) find $\mathcal{Q} = 0.95$ in G35.20−0.74N, suggesting a centrally condensed distribution of protostars. Busquet et al. (2019) obtain $\mathcal{Q} = 0.68$ for GGD 27, suggesting weak subclustering. Compared with G35.20−0.74N and GGD 27, the three CMZ protoclusters display a higher level of hierarchical subclustering.

### 4.3. *Lack of massive fragments in the three protoclusters*

We compare the dust mass distribution of our sample with the counterparts of similar sizes in mini-starbursts Sgr B2 M and N, the only two protoclusters in the CMZ that have been imaged with ALMA at 1.3 mm with a similar linear resolution (500 AU) to ours (Budaiev et al. 2024). The upper left panel of Figure 3 shows the cumulative distributions of the dust mass in different environments. For Sgr B2 M and N, we only consider sources that are classified as "cores" by the authors within the radius of 0.5 and 0.4 pc, respectively, which are the radii of the two protoclusters (Schmiedeke et al. 2016).

---

[2] https://github.com/SeamusClarke/FragMent. We have modified the distance calculations in the MST module of *FragMent* by replacing the built-in Euclidean distances with spherical distances.

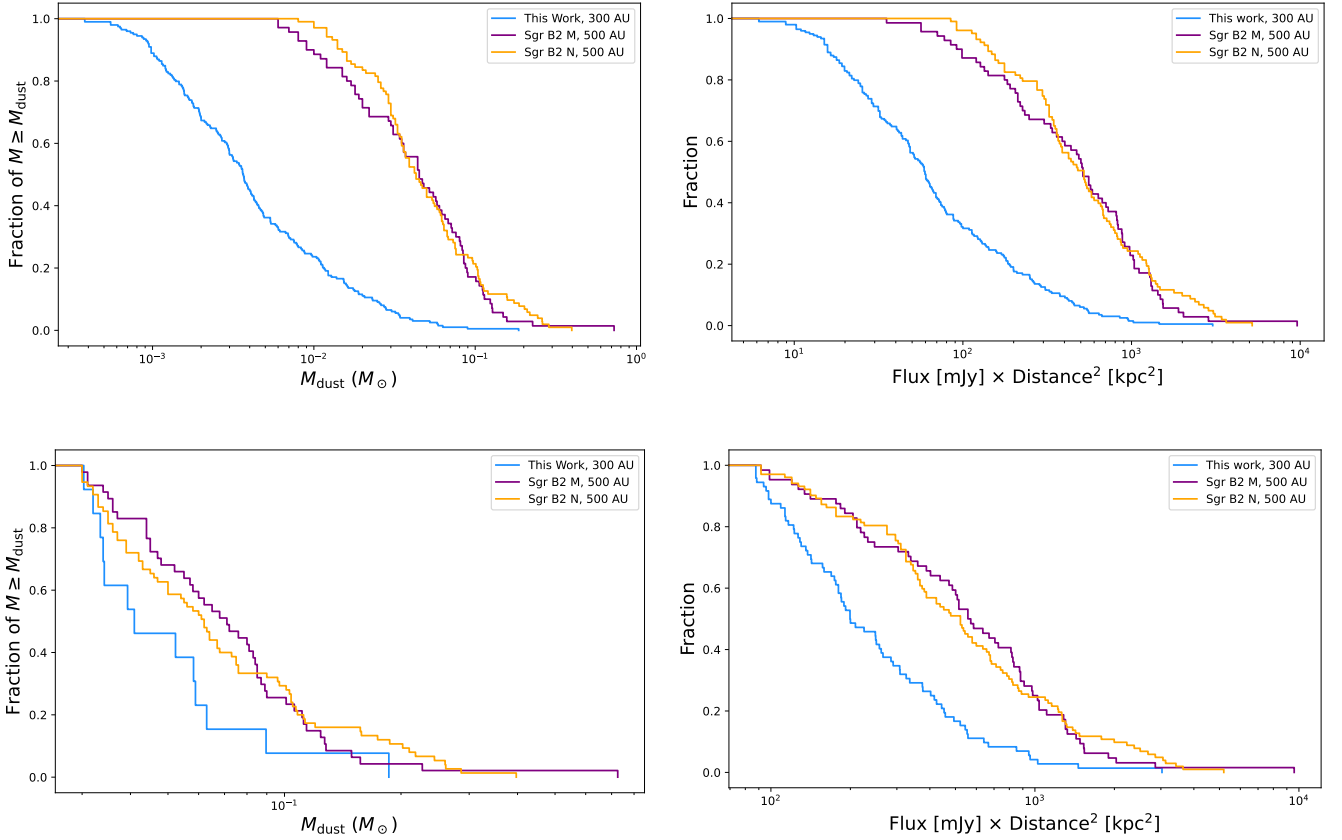

**Figure 3**: Cumulative distributions of dust mass (left column) and "luminosity" (right column; given by flux[mJy] × Distance$^2$ [kpc$^2$]) of the fragments in our sample and their counterparts of similar sizes in Sgr B2 M and N (Budaiev et al. 2024). The shaded areas indicate the ranges of errors. The upper panels show complete distributions while the bottom panels show the cropped distributions.

The dust masses of our sample are systematically lower than those in Sgr B2 M and N. To mitigate the effects of different sensitivities of these observations, we have cropped the samples to include only the sources with masses greater than the 6$\sigma$ limit of Sgr B2 N, as shown in the bottom left panel of Figure 3. Compared with Sgr B2 M and N, our sample lacks massive fragments with $M_{\rm dust} \gtrsim 0.2\ M_\odot$.

Different assumptions of dust temperature and opacity adopted in these studies could result in variation in mass up to a factor of a few. We further compare the distributions of dust luminosities for different populations, where no assumptions of dust temperature or opacity are involved. The top right panel of Figure 3 shows that the luminosities of our sample are systematically lower than that in mini-starbursts. The systematic shift in the luminosity distribution might be a combined result of different sensitivities and angular resolutions of these observations. Similarly, cropped distributions of luminosity are shown in the bottom right panel of Figure 3. Our sample shows a "top-light" luminosity distribution with respect to Sgr B2 M and N, suggesting that our sample has a lower fraction of luminous fragments and a lower mean luminosity.

The low luminosities could be explained by low protostellar masses. A lower mean luminosity of the fragments in our sample might indicate that they have lower protostellar masses compared to those in Sgr B2 M and N. On the other hand, the submillimeter luminosities of Class 0/I YSOs could increase as their accretion rates increase (Johnstone et al. 2013; Yoo et al. 2017; Contreras Peña et al. 2020). The protostellar accretion is likely more active in Sgr B2 M and N, as evidenced by more active outflows (e.g., Higuchi et al. 2015; Lu et al. 2021).

### 4.4. *Implications to cluster formation in the CMZ*

The higher level of subclustering of the three protoclusters is likely attributed to the stronger turbulence in the CMZ clumps. The results of N-body simulations by Goodwin & Whitworth (2004) show that a higher initial velocity dispersion of a cluster leads to a higher level of subclustering. The observed hierarchical clustering of fragments in the three CMZ protoclusters reflects the fractal structure of natal molecular clouds, which is likely a consequence of high velocity dispersions (FWHM∼5–10 km s$^{-1}$, Krieger et al. 2017) due to strong turbulence.

Assuming that each fragment forms a single star, we estimate a star formation rate (SFR) for each clump by counting fragments above the completeness mass limit and (if any) H II regions and extrapolate to the whole population by assuming the Kroupa initial mass function (IMF) (Kroupa 2001) using Equation (9) and (10) in Budaiev et al. (2024). The lower limit of the Kroupa IMF in the numerator is chosen to be 0.01 $M_\odot$, with an added lowest mass regime following Sanhueza et al. (2019); the lower limit in the denominator is chosen to be the corresponding stellar mass of the completeness mass limit of fragments, where the fragment-to-star efficiency is assumed to be 30% (Lada & Lada 2003). The number of the H II regions in each fragment cluster is taken from Table 5 of Lu et al. (2019a). The timescale of star formation is assumed to be $t_{\rm SF}$ = 0.5 Myr, which is consistent with the typical lifetime of H$_2$O masers (Breen et al. 2010) that are widely detected in these clumps (see Figure 1; Lu et al. 2019a). The SFRs are then calculated to be $2\times10^{-5}$ $M_\odot$ yr$^{-1}$ for 20 km s$^{-1}$ C1 and $2\times10^{-4}$ $M_\odot$ yr$^{-1}$ for 20 km s$^{-1}$ C4 and Sgr C C4.

Although these values should be taken as order of magnitude estimates due to uncertainties in e.g., the stellar mass and the star formation timescale (see Section A.1), it is reasonable to make SFR-related comparisons between our sample and Sgr B2 M and N, as the linear resolution of the two studies are comparable and the methods are essentially the same.

To explore the difference of star formation in our sample and in Sgr B2 M and N, we compute two quantities for each of these protoclusters: 1) the ratio of the total fragment mass $M_{\rm frag}$ over the clump mass $M_{\rm cl}$, which is defined as the star formation efficiency (SFE); 2) the ratio of the inferred stellar mass $M_{\rm inf}$, estimated by extrapolating the stellar mass above the completeness limit to the full mass spectrum, over the clump mass, which is defined as the high-mass star formation efficiency (HMSFE) for it largely relies on the presence of high-mass stars. The results are summarized in Table 2. An intriguing finding is that while the SFE is consistent within a factor of two, the HMSFE in Sgr B2 M and N is about one order of magnitude higher than our sample. This result suggests that Sgr B2 M and N are forming high-mass stars more efficiently than our sample, despite the similar efficiency of converting gas into stars in these clumps.

This phenomenon might be attributed to different cluster formation modes in these protoclusters. The $\mathcal{Q}$ parameter is estimated to be 0.75±0.03 for Sgr B2 M and 0.79±0.02 for Sgr B2 N (Budaiev et al. in prep), suggesting extremely weak or no subclustering. Compared to our sample, the $\mathcal{Q}$ values for Sgr B2 M and N are on average ∼50% higher (see Table 2). Using hydrodynamic simulations of giant molecular clouds with different initial density profiles ($\rho(r)$), Chen et al. (2021) find that a centrally concentrated protocluster produced by a steeper density profile is able to form more massive central clusters than a hierarchically clustered protocluster produced by a shallower density profile as benefiting from higher accretion rates. The authors also find that the integrated SFE remains similar regardless of the level of subclustering, and is determined by the interplay between gravity and feedback. The different levels of subclustering of our sample and Sgr B2 M and N likely reflect different cluster formation modes: hierarchical cluster formation versus centrally condensed cluster formation. The accretion rates vary drastically in different cluster formation modes, which can lead to different efficiencies of high-mass star formation.

We note that the hierarchical clustering and low SFR could also indicate the incipient star formation in the three protoclusters. As a protocluster evolves, its subclusters could merge and the SFR would increase (e.g., Vázquez-Semadeni et al. 2019). The stronger subclustering and lower HMSFE of our sample might be attributed to its younger age with respect to Sgr B2 M and N. To address this question in the future, high-resolution infrared data is required to character-ize the YSO populations in these protoclusters, which can be used to determine their evolutionary stages. Comparing this with the evolution of $\mathcal{Q}$ from numerical simulations, we will be able to examine if the different subclustering levels are affected by different evolutionary stages.

**Table 2**: Summary of protocluster properties

| Protocluster | $M_{\rm frag}$ ($M_\odot$) | $M_{\rm inf}$ ($M_\odot$) | $M_{\rm cl}$ ($M_\odot$) | SFE | HMSFE | $\mathcal{Q}$ |
|---|---|---|---|---|---|---|
| 20 km s$^{-1}$ C1 | 29 | 10 | 1320[a] | 0.02 | 0.01 | 0.58±0.02 |
| 20 km s$^{-1}$ C4 | 71 | 119 | 2420[a] | 0.03 | 0.05 | 0.56±0.02 |
| Sgr C C4 | 67 | 113 | 2051[a] | 0.03 | 0.06 | 0.45±0.01 |
| Sgr B2 M | 456[b] | 6882[b] | 9672[c] | 0.05 | 0.71 | 0.75±0.03 |
| Sgr B2 N | 705[b] | 2812[b] | 27897[c] | 0.03 | 0.10 | 0.79±0.02 |

[a] The clump masses are estimated using the total flux within the 3$\sigma$ contour of the SMA 1.3 mm continuum (Lu et al. 2019a, see Figure 1) and the average dust temperature from Battersby et al. (2024a).
[b] The $M_{\rm frag}$ is estimated using the catalogs presented in Budaiev et al. (2024) and only the sources classified as "cores" within the cluster radius are considered. The $M_{\rm inf}$ is calculated by multiplying the SFR with the star formation timescale $t_{\rm SF}$ in Budaiev et al. (2024).
[c] Adopted from Schmiedeke et al. (2016).

## 5. CONCLUSIONS

We present so far the highest-resolution ALMA 1.3 mm continuum images of three protoclusters in the Central Molecular Zone: 20 km s$^{-1}$ C1, 20 km s$^{-1}$ C4, and Sgr C C4. We extract a total of 199 compact millimeter sources using *astrodendro*, referred to fragments. These fragments represent the first sample of candidates of protostellar envelopes and disks and kernels of prestellar cores in the three protoclusters.

Compared to the protoclusters in the Galactic disk, the three protoclusters display a higher level of hierarchical clustering, likely attributed to the stronger turbulence in the CMZ clumps. Compared to the mini-starbursts in the CMZ, Sgr B2 M and N, the three protoclusters also show stronger subclustering. Moreover, the three protoclusters lack massive fragments and are less efficient in forming high-mass stars than Sgr B2 M and N by one order of magnitude, despite a similar overall efficiency of converting gas into stars. The lower high-mass star formation efficiency in the three protoclusters might be because they undergo hierarchical cluster formation in which the accretion rates are lower.

## ACKNOWLEDGMENTS

We thank the referee for providing constructive comments which substantially improved the quality of the paper. We thank Drs. Hui Li, Pak Shing Li, and Claudia Cyganowski for helpful discussions. This work is supported by the National Key R&D Program of China (No. 2022YFA1603100) and the Strategic Priority Research Program of the Chinese Academy of Sciences (CAS) Grant No. XDB0800300. X.L. acknowledges support from the National Natural Science Foundation of China (NSFC) through grant Nos. 12273090 and 12322305, the Natural Science Foundation of Shanghai (No. 23ZR1482100), and the CAS "Light of West China" Program No. xbzg-zdsys-202212. A.G. acknowledges support from the National Science Foundation from grants AAG 2008101 and 2206511 and CAREER 2142300. N.B. acknowledges support from the Space Telescope Science Institute via grant No. JWST-GO-05365.001-A. Y.C. was partially supported by a Grant-in-Aid for Scientific Research (KAKENHI number JP24K17103) of the JSPS. H.B.L. is supported by the National Science and Technology Council (NSTC) of Taiwan (Grant Nos. 111-2112-M-110-022-MY3, 113-2112-M-110-022-MY3). K.Q. acknowledges National Natural Science Foundation of China (NSFC) grant Nos. 12425304 and U1731237, and the National Key R&D Program of China with Nos. 2023YFA1608204 and 2022YFA1603103. S.F. acknowledges support from the National Science Foundation of China (12373023, 1213308). This paper makes use of the following ALMA data: ADS/JAO.ALMA#2018.1.00641.S. ALMA is a partnership of ESO (representing its member states), NSF (USA) and NINS (Japan), together with NRC (Canada), MOST and ASIAA (Taiwan), and KASI (Republic of Korea), in cooperation with the Republic of Chile. The Joint ALMA Observatory is operated by ESO, AUI/NRAO and NAOJ. Data analysis was in part carried out on the open-use data analysis computer system at the Astronomy Data Center (ADC) of NAOJ. This research has made use of NASA's Astrophysics Data System.

*Facilities:* ALMA

*Software:* CASA (CASA Team et al. 2022), APLpy (Robitaille & Bressert 2012), Astropy (The Astropy Collaboration et al. 2018; Astropy Collaboration et al. 2022), astrodendro (http://www.dendrograms.org/)

10 S. ZHANG ET AL.10 S. ZHANG ET AL.10 S. ZHANG ET AL.10 S. ZHANG ET AL.10 S. ZHANG ET AL.

# APPENDIX

## A. UNCERTAINTIES

### A.1. *Star formation rate*

As discussed in Section Section 3.2, the major uncertainties in mass estimates include the assumptions of optically thin dust and uniform dust temperature, choices of dust opacity, and potential contamination of free-free emission. The mass uncertainties will also propagate into the SFR estimates.

To investigate how the assumption of dust temperature impacts the SFR estimates, we perform a test by assuming a power-law distribution of the dust temperature versus flux density following Lu et al. (2020). The dust temperature of the fragment with the lowest flux density is assumed to be 20 K and the highest 200 K; this gives a powerlaw index of 0.37. Compared with the assumption of a constant $T_{\rm dust} = 50$ K, adopting a power-law temperature distribution results in a $\lesssim$10–30% variation in the SFR estimates.

At the small spatial scale probed by our data (330 AU), the fragment mass may not dominate the total mass of a protostar-disk system. For example, in the Keplerian disk system in Sgr C C4 (Lu et al. 2022), the fragment mass is $\lesssim$10% of the central stellar mass. In addition, we adopt a star formation timescale of $t_{\rm SF} = 0.5$ Myr (Section 4.4), which is $\sim$30% shorter than that adopted in the SFR estimates for Sgr B2 M and N ($t_{\rm SF} = 0.74$ Myr, Budaiev et al. 2024).

With these caveats in mind, we emphasize that although the SFR estimates should be taken as order of magnitude estimates, the proposed method enables a direct comparison of SFR between our sample and Sgr B2 M and N.

### A.2. *$\mathcal{Q}$ parameter*

The uncertainty associated with the $\mathcal{Q}$ parameter is estimated by calculating the $\mathcal{Q}$ parameter for the fragments of which 20% are randomly discarded and repeating this process for 1000 times. The standard deviation of the Gaussian fit to the distribution of the $\mathcal{Q}$ parameter is taken as the uncertainty.

The morphology analysis based on the $\mathcal{Q}$ parameter could be, somehow, biased by (1) the sample size and (2) the limited capability of detecting faint sources due to the sensitivity of observations and the obscuring of nearby sources. The estimates of $\mathcal{Q}$ in Zhang et al. (2022) and Busquet et al. (2019) involve $\leq$25 sources, which may not be as robust as ours (with $\sim$50–100 sources). Using method described above, we estimate an uncertainty of $\sim$0.02 of the $\mathcal{Q}$ value for each of the two regions in Galactic disk. In addition, our continuum sensitivity is comparable with that of Busquet et al. (2019) but 3–4 times poorer than Zhang et al. (2022) (after normalized to the same distance). It is likely that we have missed more faint sources which are below our detection thresholds. We perform simple tests by calculating the $\mathcal{Q}$ parameter for our sample that is cropped by increasing thresholds and find that for a known hierarchical cluster, missing faint sources could lead to either a higher or lower $\mathcal{Q}$ value. Also, faint sources in clustered environments can be obscured due to crowding and/or by nearby luminous sources (Ascenso et al. 2009). These effects are difficult to quantify without the knowledge of the cluster configuration, demanding analytical models and numerical simulations for further examination.